\input hyperlatex
\documentstyle[aps,prb,floats,epsf]{revtex}
\begin{document}
\twocolumn[\hsize\textwidth\columnwidth\hsize\csname@twocolumnfalse%
\endcsname
\title{Thermally fluctuating superconductors in two dimensions}
\author{\href{http://sachdev.physics.yale.edu/Sachdev.html}{Subir Sachdev}
and Oleg A. Starykh}
\address{Department of Physics, Yale University\\
P.O. Box 208120,
New Haven, CT 06520-8120, USA}
\date{January 1, 2000}
\maketitle
\begin{abstract}
We describe the different regimes of finite temperature dynamics
in the vicinity of a zero temperature superconductor to insulator quantum phase
transition in two dimensions. New results are obtained for a low
temperature phase-only hydrodynamics, and for the intermediate
temperature quantum-critical region. In the latter case, we
obtain a universal relationship between the frequency-dependence
of the conductivity and the value of the d.c. resistance.
\end{abstract}
\pacs{PACS numbers:}
]

In many interesting two-dimensional superconducting systems
\cite{goldman,clarke,joeo}, such as Josephson junction arrays,
granular superconducting films, and the high temperature
superconductors, it appears that the electrons bind into Cooper
pairs below a pairing temperature ($T_P$) that is well above the
Kosterlitz-Thouless transition, at $T_{KT}$, to long-range
superconducting order~\cite{uemura,doniach,fgg,nandini,emery}. It
is natural to search for a direction in parameter space where
$T_{KT}$ vanishes at a $T=0$ superconductor-insulator quantum
phase transition \cite{sondhi,book}, and to then expand in the
deviation from the quantum critical point---it is not necessary
for this point to be experimentally accessible for such an
approach to be valuable, as it offers a controlled description of
an intermediate coupling regime. We describe crossovers in the
dynamics near the critical point: new results are obtained for a
low temperature phase-only regime, and for the intermediate
temperature `quantum-critical' region. In the latter
regime we describe the frequency ($\omega$) dependent conductivity
($\sigma$) in terms of a single dimensionless parameter, $\gamma
(T)$, which determines the d.c. conductivity; $\gamma (T)$ can
also be determined by separate static measurements.

\begin{figure}[ht]
\epsfxsize=2.4in
\centerline{\epsffile{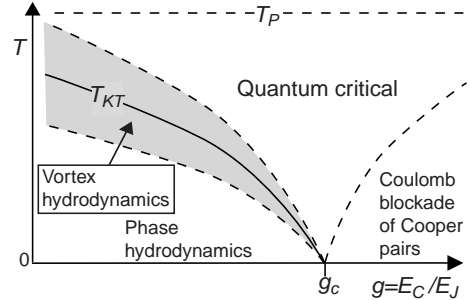}}
\caption{Phase diagram of $H$ in (\protect\ref{hd}).
Dashed lines are crossover, while the full
line is a phase transition. $\rho_s (T)$ is non-zero for
$g<g_c$, $T< T_{KT}$; at $T=0$
$\rho_s (0) \sim (g_c - g)^{z \nu}$ with $z$ the dynamic exponent
and $\nu$ the correlation length exponent ($z=1$ for $H$).
The ground state for $g>g_c$ is an insulator. If one studies $H$
by a tunneling probe to a good superconductor, there will be a zero
voltage Josephson current in the region where $\rho_s (T)$ is non-zero, while
for $g > g_c$ and low $T$, a Cooper pair
current will only flow under a finite voltage which
overcomes the Coulomb blockade.}
\label{figp}
\end{figure}

For clarity, we will present our results in the context of
a familiar microscopic model for the superconductor-insulator
transition; however, our results generalize to a much wider class
of systems, and this will be discussed towards the end of the paper.
We consider a array of superconducting quantum
dots at the sites, $i$, of a regular two-dimensional lattice. The
operator $\hat{n}_i$ measures the number of electron pairs on dot
$i$, and $\hat{\varphi}_i$ is its canonically conjugate phase
($[\hat{n}_i , \hat{\varphi}_j] = \delta_{ij}$). We consider the
Hamiltonian \cite{d1}
\begin{equation}
H = (E_C /2) \sum_i (\hat{n}_i - N_0)^2 - E_J \sum_{<ij>} \cos(
\hat{\varphi}_i - \hat{\varphi}_j ),
\label{hd}
\end{equation}
where $N_0$ is the mean integer number of Cooper pairs on each
dot, $<ij>$ represents nearest neighbor pairs, $E_C$ is the
charging energy of a dot, and $E_J$ is the Josephson tunnel
coupling between dots. This model exhibits a superconductor to
insulator transition as the dimensionless parameter $g=E_C/E_J$ is
increased through a critical value $g_c$ - the phase diagram in
the $T$, $g$ plane is summarized in Fig~\ref{figp}.
We will be
mainly concerned with the region $g \leq g_c$.
For $T<T_{KT}$, in the superconducting phase,
the conductivity has the
delta-function contribution $\mbox{Re}[\sigma] = (e^{\ast
2} \pi \rho_s (T)/\hbar^2) \delta ( \omega )$, ($e^{\ast} = 2 e$),
which defines the
superfluid stiffness $\rho_s (T)$ (in units of energy).
The hyperscaling property of the
quantum critical point implies that {\em all} dynamic properties in its
vicinity are entirely characterized by a single energy, $\rho_s (0)$,
which vanishes at $g=g_c$.
In particular,
the conductivity obeys\cite{book}
\begin{equation}
\sigma ( \omega, T) = \frac{e^{\ast 2}}{h} \Sigma \left(
\frac{\hbar \omega}{k_B T}, \frac{\rho_s (0)}{k_B T} \right)
\label{e1}
\end{equation}
where $\Sigma$ is a completely universal scaling function. Such a
two argument scaling form is not overly constraining, and permits a
rich variety of behavior which we shall describe here as a function of the second
argument $\lambda \equiv \rho_s (0)/k_B T$.
Earlier studies of superfluid dynamics (made without reference to quantum phase
transitions) emerge in limiting regimes, with (\ref{e1}) only placing
restrictions on certain parameter values.

Demanding
consistency of (\ref{e1}) with the definition of $\rho_s (T)$ above
immediately leads to
interesting conclusions\cite{ssrelax}: ({\em i\/}) $\rho_s (T) /\rho_s (0)$ is a
universal function of $\lambda$; ({\em ii\/}) the Kosterlitz
Thouless transition occurs at the universal value $\lambda =
\lambda_c$ where $\Sigma$ is singular; ({\em iii\/})
$T_{KT} = \rho_s (0) / \lambda_c$. Aspects of the
data on the high temperature
superconductors are consistent with such trends\cite{uemura,emery},
suggesting the proximity of an insulator of Cooper pairs (with stripe order).

We now describe the evolution of dynamic properties with
decreasing $\lambda$.

\noindent
\underline{Phase hydrodynamics:}
At very large $\lambda$ ($k_B T \ll
\rho_s (0)$) vortex excitations are exponentially suppressed, and
we can derive an effective quantum action,
$S_{\varphi}$ for the continuum phase
variable $\varphi (x, \tau)$ (where $x$ is a two-dimensional
spatial co-ordinate and $\tau$ is imaginary time) in a gradient
expansion; the resulting action is valid only at the
longest scales, and is in the spirit of the
`chiral lagrangians' of particle physics:
\begin{displaymath}
S_{\varphi} = \int \!\! d^2 x d \tau \!\! \left[
\frac{\rho_s (0)}{2} \left( \partial_{\mu} \varphi \right)^2 -
\frac{A_1 (\hbar v)^2}{2 \rho_s (0)} \left( \partial_{\mu} \varphi
\right)^2 \left( \partial_{\nu} \varphi
\right)^2 \right].
\end{displaymath}
Here $\mu,\nu$ are spacetime indices with $\partial_{\mu} =
(\nabla_x, \partial_{\tau}/v)$, and $v$ is the velocity of the
`spin-wave' excitations produced by the harmonic terms in
$S_{\varphi}$ ($v$ remains non-singular through the quantum-critical point).
The non-linear term, arises from integrating out amplitude fluctuations,
and leads to spin-wave scattering; consistency of the resulting transport
properties with (\ref{e1}) demands that the scattering
cross-section be universally determined by $\rho_s (0)$--hence the
dimensionless number $A_1$ is {\em universal}. In general spatial
dimension, $d$, $A_1$ multiplies $(\hbar v)^{2/(d-1)} [\rho_s (0)]^{(d-3)/(d-1)}$,
from which it is evident that the universality of $A_1$ holds only
for $1<d<3$; we computed $A_1$
for $H$ in an expansion\cite{ssrelax} in $\epsilon=3-d$ and obtained
\begin{displaymath}
A_1 =  \left[\frac{10 (4 \pi)^{-(d+1)/2}}{\Gamma
((d+1)/2) \epsilon }\right]^{2/(d-1)} \left( 1 - \frac{11 \epsilon}{30} +
O(\epsilon^2) \right).
\end{displaymath}
Determination of the $T>0$ transport properties of $S_{\varphi}$ requires
solution of the appropriate quantum kinetic equations--we will not
do this here. The procedure is closely analogous to
early work\cite{khalat} on phonon transport in Galilean-invariant superfluids;
an important difference is that these systems had a
cubic non-linearity which is forbidden in our case by
particle-hole symmetry.

\noindent
\underline{Vortex hydrodynamics:}
In the shaded region of
Fig~\ref{figp}, in the vicinity of $T_{KT}$, the vortices
proliferate, and the dynamics can
be described by a well-developed classical theory\cite{ahns}
for the vortices alone.
This theory is contained within $\Sigma$ for $\lambda \approx
\lambda_c$, and compatibility constrains various prefactors.
So {\em e.g.} for $T>T_{KT}$ the response of the free vortices is
controlled by their diffusivity, $A_2 \hbar v^2 /\rho_s (0)$, and
their screening length, $A_3 (\hbar v/\rho_s (0))
e^{A_4/(\lambda_c-\lambda)^{1/2}}$; in general the $A_{2-4}$ are
arbitrary dimensionless scale factors---however, near $g=g_c$ they
become universal numbers.

\noindent
\underline{Quantum critical:} Discussion of this small $\lambda$ regime will
occupy the remainder of the paper.
Previous work~\cite{kedar,igor,ssqhe,book}
relied upon expansions in
either small $3-d$, or
small $d-1$, or large $N$ (the number of real order parameter components).
Here we shall present a theory directly for the physical case
$N=2$, $d=2$.

Unlike the two previous regimes discussed above, it is no longer
possible to decouple the spin-wave and vortex degrees of freedom.
We will therefore use the complex superconducting order parameter
$\psi (x,t)$ ($t$ is real time) which is the continuum limit of
the lattice operator $e^{i \hat{\varphi}_i}$, and allow for both
amplitude and phase fluctuations in $\psi$. Our theory follows
from two hypotheses: ({\em i\/})  The
equal-time correlations of $\psi$ are controlled by a Gaussian effective
action. Evidence for this hypothesis emerged in detailed studies
of order parameter correlations in the quantum-critical
region\cite{CSY}--the non-Gaussian components of the $\psi$
correlations are weak because the order parameter anomalous
dimension at the quantum-critical
point, $\eta \approx 0.03$, is so small.
(This small $\eta$ also shows why $\psi$ is
preferred over the `dual order parameter' measuring vorticity
\cite{dual}--the latter has an appreciable anomalous dimension
\cite{sudbo}.)
({\em ii\/}) The time evolution of $\psi (x,t)$
is described by classical equations of motion. The characteristic
relaxation (phase coherence)
time in the quantum-critical regime is of order\cite{SY} $\hbar
/k_B T$; however the dominant spectral weight is at frequencies
$\hbar \omega < k_B T$, and there is good
evidence\cite{book,ssrelax} that the errors made by focusing on
this low-frequency classical regime are quite small. We note that
a different classical dynamic model was considered in Ref.~\onlinecite{ssrelax},
but it does not apply to transport properties.

We now define our model for quantum-critical transport, and then
present its exact (numerical) solution. In additional to the field
$\psi (x,t)$, we will need the canonically conjugate variable
measuring density fluctuations, $\delta n (x,t)$, which is the
continuum limit of $\hat{n}_i - N_0$.
The equal-time correlations of $\psi$ and $\delta n$ are described
by the partition function
\begin{eqnarray}
{\cal Z} &=& \int {\cal D} \psi (x) {\cal D} \delta n(x) \exp (-
({\cal H}_1 + {\cal H}_2)/k_B T ),
\label{part} \\
{\cal H}_1 &=& \int d^2 x \left[ |\nabla \psi|^2 + \xi^{-2} (T)
|\psi |^2 \right],
\label{h1} \\
{\cal H}_2 &=& \left(1/[2 \chi_u (T)]\right) \int d^2 x (\delta
n(x))^2 .
\label{h2sr}
\end{eqnarray}
The overall scale of $\psi$ is arbitrary,
and has been adjusted to obtain a unit coefficient for the
gradient term in (\ref{h1}).
The correlation length, $\xi (T)$, (assumed to be
larger than all microscopic scales), and the
compressibility $\chi_u (T)$, are determined by the underlying
quantum physics, and will generally be regarded as unknown
input parameters in our dynamic theory; proximity to the
quantum-critical point does (weakly) restrict their allowed
$T$ dependencies, but we can evisage a more general application of
our dynamic theory, without reference to quantum criticality, in
which case $\xi (T)$, $\chi_u (T)$ are arbitrary.
The time evolution of $\psi$, $\delta n$, is defined by
a minimal model---the Hamilton-Jacobi equations of ${\cal H}_1
+ {\cal H}_2$, and the only non-zero Poisson bracket
\begin{equation}
\left\{ \delta n(x), \psi (x') \right\}_{\rm P.B.} =
(i/\hbar) \psi (x) \delta^2 (x-x'),
\label{pb}
\end{equation}
which is the continuum, classical limit of the commutatator
between $\hat{\varphi}_i$ and $\hat{n}_i$.
Notice that $\hbar$ appears on the r.h.s. even though we are
considering classical equations.
The equations of motion implied by (\ref{h1})-(\ref{pb}) are simple
and familiar. They are the continuity equation
\begin{equation}
\partial \delta n(x,t)/\partial t + \nabla \cdot J(x,t) = 0,
\label{cont}
\end{equation}
where $J = -(i/\hbar)(\psi^{\ast} \nabla \psi - \psi \nabla \psi^{\ast})$,
and
\begin{equation}
\partial \psi(x,t)/\partial t = -(i/\hbar) \Phi(x,t) \psi(x,t),
\label{josephson}
\end{equation}
which is the Josephson equation, with the
electrochemical potential $\Phi(x,t) = \delta n(x,t)/\chi_u (T)$.
We restate our dynamical model for quantum critical
transport: choose a set of equal-time
initial conditions for $\psi$ and $\delta n$ from the thermal ensemble
defined by ${\cal Z}$, and evolve them deterministically
by (\ref{cont}) and (\ref{josephson}). Correlation functions are
determined by the average over initial
conditions.

We will shortly present convincing numerical evidence that
(\ref{part})-(\ref{josephson}) define a sensible continuum theory
free of short distance or short time (`ultraviolet') divergences;
this is supported by perturbative and renormalization group arguments,
which we do not describe here.
So unlike $S_{\varphi}$, the present theory
describes the couplings of vortex and
spin-wave fluctuations at different length scales. It is helpful to
visualize the continuum theory by coarse-graining to a lattice
spacing, $a$: the shorter distance degrees of freedom lead to
fluctuations in the phase and amplitude of $\psi$, but
the long distance transport properties are insensitive
to the value of $a$. The absence of factors of $a$ in the final results allows us
to deduce their functional dependence on
$\xi (T)$, $\chi_u (T)$ by
simple engineering dimensional analysis.
In this manner we conclude that
\begin{equation}
\sigma (\omega, T) = \frac{e^{\ast 2}}{h} \gamma (T) \Sigma_c \left(
\gamma (T) \frac{\hbar \omega}{k_B T} \right),
\label{cond}
\end{equation}
where $\Sigma_c$ is a universal function (`exact' numerical
results are below), and the dimensionless  $\gamma (T)$ is defined by
\begin{equation}
\gamma (T) \equiv [k_B T \chi_u (T) \xi^2(T)]^{1/2}.
\end{equation}
Consistency of (\ref{cond}) with (\ref{e1}) only requires that
$\gamma (T)$ is a universal function of $\lambda$ (also recall
that the present quantum-critical theory is valid only for small
$\lambda$).
The result (\ref{cond}) has a clear experimental
signature: it implies a correlation between the value
of the d.c. conductivity and the inverse frequency-width of $\sigma
(\omega)$, as they are both determined by $\gamma (T)$.
More stringent comparisons can be made by using the measured d.c.
conductivity to determine the unknown $\gamma (T)$, and then using
our numerical results below for $\Sigma_c$ to determine
the $\omega$ dependence of $\sigma$. It should be noted
that our present computations for $\Sigma_c$ will not apply for
very large $\hbar \omega/k_B T$, for then a full quantum theory
is necessary, and we crossover to the phase-coherent regime
discussed earlier~\cite{fgg,kedar}.

Exactly at $g=g_c$, hyperscaling arguments imply that
$\xi (T) \sim T^{-1/z}$ and $\chi_u (T) \sim T^{(2-z)/z}$; so $\gamma (T)$
is $T$-independent for any $z$, and is expected to be a universal number.
For the model $H$ in (\ref{hd}), this universal number can be computed\cite{CSY}
in a $1/N$ expansion:
\begin{equation}
\left.\gamma (T) \right|_{g=g_c} \!= \!\left[ \frac{\sqrt{5}}{4 \pi
\ln((\sqrt{5}+1)/2)} \right]^{1/2}\!\!\!\! \left(1 - \frac{0.5468}{N} \right).
\label{gsr}
\end{equation}

We turn to our numerical results for $\Sigma_c$. The
simulations were carried out on an $N\times N$ square
lattice of spacing $a$
with periodic boundary conditions, and $a \ll \xi(T) \ll N a$.
For each $a$, successively
larger values of $N$ were used, until the results became
$N$-independent, and the continuum limit was then approached
as $a \rightarrow 0$.
The lattice form of ${\cal H}_1 + {\cal H}_2$ was obtained by mapping the
momentum ($k_x$, $k_y$) dependence of the couplings of the continuum Hamiltonian
under $k_x^2 \rightarrow (4/a^2)[ K_1 \sin^2 (k_x a/2)
+ K_2 \sin^2 (k_x a) + K_3 \sin^2 (3 k_x a/2)]$, and similarly for
$k_y$ (for $|\nabla \psi|^2$, this amounts to including first,
second, and third neighbor couplings between the $\psi$). We chose
two different sets of values for $K_{1,2,3}$ to test the
independence of the continuum theory on the lattice
realization---(A) the familiar $K_1=1$, $K_2=K_3=0$, and
(B) $K_1 = 3/2$, $K_2 = -3/20$, $K_3 = 1/90$ for which
$k_x^2 \rightarrow k_x^2 ( 1 + O (k_x a)^6 )$.
The initial conditions specified by ${\cal Z}$ form a Gaussian
ensemble, and are easily generated in a single sweep. The time evolution was
carried out by a fourth order predictor-corrector algorithm with a
time step determined to conserve total energy to a
relative accuracy better than $10^{-5}$.
We measured the autocorrelation
function of the total current, $C_J$, by averaging over 3000 initial
conditions; $C_J$ is normalized such that
\begin{equation}
\Sigma_c (\overline{\omega}) = \int_0^{\infty} \! d \overline{t}
\, C_J (\overline{t}) \, e^{i \overline{\omega} \, \overline{t}},
\label{ft}
\end{equation}
where $\overline{\omega} = \gamma (T) \hbar \omega/(k_B T)$
and $\overline{t} = k_B T t/(\hbar \gamma (T))$.

\begin{figure}[ht]
\epsfxsize=3.2in
\centerline{\epsffile{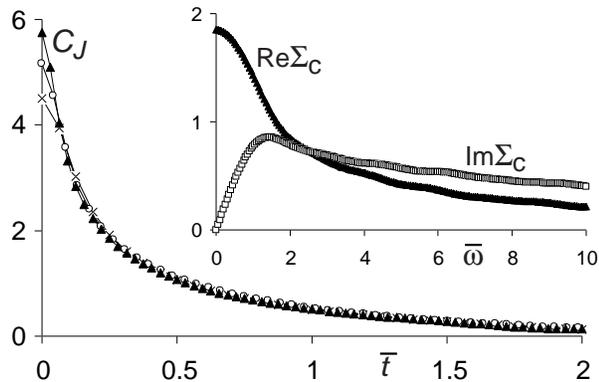}}
\caption{Current autocorrelation, $C_J$, of
(\protect\ref{part})-(\protect\ref{josephson}),
with lattice realization A, $N=64$, and
$a/\xi(T)= 1/8$ (crosses), $\protect\sqrt{2}/16$ (open circles), $1/16$ (triangles).
Inset: real and imaginary parts of $\Sigma_c$,
obtained from (\protect\ref{ft})
for $a/\xi(T) = 1/16$;  for large $|\overline{\omega}|$ we have
$\mbox{Re} \Sigma_c \sim 1/|\overline{\omega}|$ and
$\mbox{Im} \Sigma_c \sim
\ln(|\overline{\omega}|)/\overline{\omega}$.
}
\label{fig1}
\end{figure}

Our results for $C_J$ and $\Sigma_c$ are in
Fig~\ref{fig1}.
Notice the
$a$ dependence at $\overline{t} = 0$---this is expected as
an analytic calculation of equal-time correlations shows
that $C_J (0) = (1/\pi) \ln (1/a) + \ldots$. However, the $a$
dependence
quickly disappears at small non-zero $\overline{t}$, and then
a universal continuum value obtains; this is strong evidence
for the existence of the continuum theory (\ref{part})-(\ref{josephson}).
We obtained $\Sigma_c (0) \approx 1.85$;
for the model $H$ in (\ref{hd}),
this combines with (\ref{gsr}) to yield
$\sigma(0,T) = 0.82 (e^{\ast 2} /h)$ at $g=g_c$.

The experiments of Rimberg {\em et al.}\cite{clarke} appear to be a
convenient testing ground for our theory. The metallic gate
screens the long-range Coulomb interactions and so justifies\cite{ziman}
the short-range coupling in ${\cal H}_2$. However, it also
introduces dissipation, which will induce additional
terms in the equations of motion {\em e.g.} (\ref{josephson}) is
modified to
\begin{equation}
\partial \psi/\partial t = -\Gamma \delta {\cal H}_1/\delta
\psi^{\ast}
-(i/\hbar) \Phi \psi + \zeta,
\label{josephsond}
\end{equation}
where $\Gamma$ is a damping co-efficient and $\zeta$ is the
associated random noise. Fortunately, it can be shown,
as in Refs~\onlinecite{kawa2},
that for large $\xi(T)$,
such modifications leave the dynamic function $\Sigma_c$
{\em unchanged}. The damping can be a relevant perturbation on the
$T=0$ quantum critical point\cite{wagen}, but this manifested
only via changes in $\xi(T)$, $\chi_u (T)$, and
$\gamma (T)$. (Outside the quantum critical region, the change in
quantum universality leaves the form (\ref{e1}) unchanged, but
with a new $\Sigma$; the structure of the phase hydrodynamics, $S_{\varphi}$ can
also be modified. Similar comments apply to the following
paragraph.)

As a second modification, we consider the inclusion of long-range
Coulomb interactions. This will modify ${\cal H}_2$ to
$ e^{\ast 2}\int d^2 x d^2 x' \delta n(x) \delta
n(x')/2|x-x'|$, and leave (\ref{cont},\ref{josephson}) unchanged,
but with $\Phi (x,t) = e^{\ast 2} \int d^2 x' \delta
n(x',t)/|x-x'|$. Assuming the existence of the continuum limit of
the classical dynamic theory, we again obtain (\ref{cond}),
but with $\gamma (T) = [k_B T \xi(T)/e^{\ast 2}]^{1/2}$; the
functional form of $\Sigma_c$ will of course be changed and
requires a separate numerical simulation. Note that at $g=g_c$,
$\gamma (T)$ is a $T$-independent universal number only if $z=1$,
the value expected for long-range interactions\cite{fgg}.

The experiments of Corson {\em et al.}\cite{joeo} on an underdoped
high temperature superconductor observe scaling closely related
to (\ref{cond}), with the frequency scale proportional to the
resistivity scale, and a scaling function with large $\omega$ behavior
similar to that in Fig~\ref{fig1}. However their fits use a
prefactor $T_{\theta}^0$ whose weak $T$-dependence disagrees with
our continuum two-dimensional theory. We speculate that
this can be explained by corrections to scaling which are appreciable
because both modifications discussed above are
present here -- damping from fermionic excitations and long-range
Coulomb interactions. Corson {\em et al.}\cite{joeo}
motivated the scaling using
the vortex theory\cite{ahns}, but did not consider bound vortex
pairs, whose contributions do not obey their scaling
assumptions.

In summary, this paper has delineated the distinct dynamic regimes
of thermally fluctuating superconductors: the low $T$
phase-only hydrodynamics, the classical vortex hydrodynamics in the
vicinity of $T_{KT}$, and the quantum-critical region where phase
and vortex fluctuations strongly coupled---here we proposed a
dynamical model in which non-linear `mode-coupling' terms demanded by the
Poisson bracket (\ref{pb}) dominate the universal, low-frequency,
dissipative dynamics.
Our approach could be extended to other
two-dimensional systems: ({\em i\/}) the magnetic field-tuned
superconductor-insulator transition, where a coupling to the
external field would be required in (\ref{h1}), and ({\em ii\/})
quantum Hall transitions.

We thank J.~Orenstein, D.~Grempel, and A.~Sudb{\o} for useful discussions.
This research was supported by NSF Grant No DMR 96--23181.



\end{document}